\documentclass[showpacs,prl,aps,twocolumn]{revtex4}

 \usepackage{bm}
 \usepackage{amsmath}
 \usepackage{amssymb}
 \usepackage{latexsym} 
 \usepackage{amsfonts} 
 \usepackage{epsfig}
 \usepackage{psfrag} 

 \newcommand{\ket}[1]{\left|#1\right>} 
 \newcommand{\bra}[1]{\left<#1\right|} 
  
 \newcommand{\nn}{\nonumber\\} 
  
 \newcommand{\f}[1]{\mbox{\boldmath$#1$}}
 \newcommand{\fk}[1]{\mbox{\boldmath$\scriptstyle#1$}}
 \newcommand{\vau}{\mbox{\boldmath$v$}}
 \newcommand{\na}{\mbox{\boldmath$\nabla$}}
 \newcommand{\bea}{\begin{eqnarray}}
 \newcommand{\ea}{\end{eqnarray}}
 \newcommand{\eea}{\end{eqnarray}}
 \newcommand{\ord}{{\cal O}}

\begin{document}
  
\title{On the detectability of quantum radiation in Bose-Einstein condensates} 

\author{Ralf Sch\"utzhold}

%\email{schuetz@theory.phy.tu-dresden.de}

\affiliation{Institut f\"ur Theoretische Physik, 
Technische Universit\"at Dresden, D-01062 Dresden, Germany}

\begin{abstract}
Based on doubly detuned Raman transitions between (meta) stable atomic
or molecular states and recently developed atom counting techniques, a
detection scheme for sound waves in dilute Bose-Einstein condensates is
proposed whose accuracy might reach down to the level of a few or even
single phonons. 
This scheme could open up a new range of applications including the
experimental observation of quantum radiation phenomena such as the
Hawking effect in sonic black-hole analogues or the acoustic analogue
of cosmological particle creation. 
\end{abstract}

\pacs{
04.70.Dy, % Quantum aspects of black holes, evaporation,
	  % thermodynamics 
03.75.Kk, % Dynamic properties of condensates; collective and
       	  % hydrodynamic excitations, superfluid flow
42.65.Dr. % Stimulated Raman scattering; 
}

\maketitle

%%%%%%%%%%%%%%%%%%%%%%%%%%%%%%%%%%%%%%%%%%%%%%%%%%%%%%%%%%%%%%%%%%%%%%%%%%%%%%%
%%%%%%%%%%%%%%%%%%%%%%%%%%%%%%%%%%%%%%%%%%%%%%%%%%%%%%%%%%%%%%%%%%%%%%%%%%%%%%%
%{\em Introduction}\quad
%%%%%%%%%%%%%%%%%%%%%%%%%%%%%%%%%%%%%%%%%%%%%%%%%%%%%%%%%%%%%%%%%%%%%%%%%%%%%%%
%%%%%%%%%%%%%%%%%%%%%%%%%%%%%%%%%%%%%%%%%%%%%%%%%%%%%%%%%%%%%%%%%%%%%%%%%%%%%%%
%
Gaseous atomic or molecular Bose-Einstein condensates \cite{bec} 
are in several ways superior to other superfluids:
Apart from a very good theoretical understanding via the mean-field
formalism (in the dilute-gas limit), they offer unprecedented options
for experimental manipulation and control.
It is possible to influence the shape, density, flow profile, and
coupling strength of Bose-Einstein condensates via external
electromagnetic fields.
Finally, these condensed gases are rather robust against the
impact of the environment such that one may reach extremely low
temperatures. 

In view of all these advantages, the question naturally arises whether  
it could be possible to measure so far unobserved quantum radiation
phenomena in a suitable set-up.
These exotic quantum effects include cosmological particle creation 
(due to the amplification of quantum fluctuations in an
expanding/contracting universe \cite{inflation,visser}) 
as well as the acoustic analogue of Hawking radiation
\cite{hawking,birrell} in ``dumb holes'' \cite{unruh,garay,visser}.

For wavelengths which are much longer than the healing length~$\xi$,
the propagation of phonons in Bose-Einstein condensates is analogous
to a scalar field in a curved space-time described by the effective
metric \cite{unruh}
\bea
\label{metric}
g^{\mu\nu}_{\rm eff}
=
\frac{1}{\varrho_0c_{\rm s}}
\left(
\begin{array}{cc}
1 & \vau_0 \\
\vau_0 & \vau_0\otimes\vau_0 - c^2_{\rm s}\bf{1}
\end{array}
\right)
\,,
\ea
which is determined by the density $\varrho_0$ and velocity $\vau_0$
of the background fluid.
For example, assuming an effectively one-dimensional stationary flow,
the point where the fluid velocity $v_0$ exceeds the local speed of
sound $c_{\rm s}$ corresponds to the sonic analogue of the horizon of
a black hole.  
The corresponding Hawking temperature is determined by the velocity
gradient \cite{unruh} 
\bea
T_{\rm Hawking}
=
\frac{\hbar}{2\pi\,k_{\rm B}}\,
\left|\frac{\partial}{\partial r}\left(v_0-c_{\rm s}\right)\right|
\,,
\ea
i.e., the characteristic length scale $\lambda$ over which the flow
changes.
Since this length scale should be large compared to the healing
length~$\xi$ (typically of order micrometer) for the curved space-time
analogy to apply, a speed of sound of order mm/s leads to an upper
bound for the typical energy of the Hawking phonons of order
$10^{-13}$~eV corresponding to a temperature on the nano-Kelvin
level. 

Moreover, since the fluid velocity equals the sound speed at the
acoustic horizon, only a limited number of these low-energy phonons
will be created by the Hawking effect -- unless one has a very large
reservoir for the condensate flow:
Since the Hawking radiation is thermal, the typical distance between
two emitted Hawking phonons is given by their characteristic
wavelength~$\lambda$ and hence it is much larger than the healing
length~$\xi$.
In addition, Bose-Einstein condensates are formed by atoms 
(or molecules) whose inter-particle distance~$a_d$ is far bigger than
the $s$-wave scattering length~$a_s$ (dilute-gas limit).
As a result, a healing length~$\xi\propto a_d\sqrt{a_d/a_s}$ contains
many atoms $\xi\gg a_d$, i.e., we have a hierarchy of length scales
$\lambda \gg \xi \gg a_d \gg a_s$.  
Consequently, the number of Hawking phonons is extremely small
compared to the number of atoms in the condensate $a_d/\lambda\lll1$.  

Similar arguments apply to the analogue of cosmological particle
creation, which require a non-stationary set-up. 
Considering the effective metric in Eq.~(\ref{metric}), there are
basically two possibilities for simulating the cosmic expansion in
Bose-Einstein condensates: 
an expansion of the condensate or a temporal variation of the speed of
sound (which can be achieved via varying $a_s$ by means of a Feshbach
resonance, for example). 
For simplicity, we shall focus on the second possibility in the
following, but the general ideas apply to both scenarios. 
The typical wavelength~$\lambda$ of the created phonons is
determined by the rate of change $\lambda=\ord(c_s^2/\dot c_s)$  
of~$c_s$ and should again be large compared to the healing
length~$\xi$ for the curved space-time analogy to apply.
In the absence of amplification mechanisms such as resonances, the
number of created phonon per wavelength is again of order one
\cite{birrell}. 

For a small number of phonons with an energy of order $10^{-13}$~eV,
the usual detection mechanisms for sound via mechanical motion are
extremely difficult to apply since the kinetic energy of a single atom 
with a velocity of order mm/s already yields this amount.
Usually, these measurements involve many atoms and thus many phonons
(limit of classical waves). 
For example, it was possible to excite phonon modes via light
scattering \cite{phonon} and to map out the dispersion relation etc.
An indirect observation of the phonon number was achieved in
ultra-sensitive temperature measurements \cite{temperature}, 
which reached very low energies. 

Fortunately, a non-mechanical detection mechanism may circumvent these
obstacles. 
For example, the ion-trap quantum computer in 
%the scheme for quantum computing in ion traps in
\cite{ions} is based on optically induced transitions which require
the simultaneous absorption of a phonon in a given mode 
(due to the detuning of the Laser).
Using the occurrence of the transition and the photon emitted during
the decay back to the ground state as an indicator for the existence
of the phonon yields an energy amplification over many orders of
magnitude; and the detection of single photons in the optical range is
difficult but feasible (in principle).
A further amplification is possible if the phonon-assisted transition
mediates between (meta) stable atomic states which can be separated or
addressed individually:
The detection of a small number of atoms can be achieved via
fluorescence measurements involving many photons
\cite{counting,raizen}. 

%%%%%%%%%%%%%%%%%%%%%%%%%%%%%%%%%%%%%%%%%%%%%%%%%%%%%%%%%%%%%%%%%%%%%%%%%%%%%%%
\begin{figure}[ht]
\includegraphics[height=3cm]{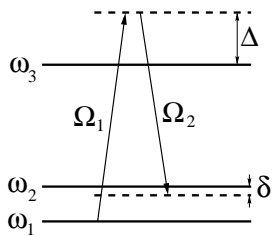}
\caption{\label{level} Sketch (not to scale) of the three-level
  ($\Lambda$) system and the doubly detuned Raman transitions denoted
  by $\Omega_{1,2}$.} 
\end{figure}
%%%%%%%%%%%%%%%%%%%%%%%%%%%%%%%%%%%%%%%%%%%%%%%%%%%%%%%%%%%%%%%%%%%%%%%%%%%%%%%

%%%%%%%%%%%%%%%%%%%%%%%%%%%%%%%%%%%%%%%%%%%%%%%%%%%%%%%%%%%%%%%%%%%%%%%%%%%%%%%
%%%%%%%%%%%%%%%%%%%%%%%%%%%%%%%%%%%%%%%%%%%%%%%%%%%%%%%%%%%%%%%%%%%%%%%%%%%%%%%
%{\em Doubly detuned Raman transitions}\quad
%%%%%%%%%%%%%%%%%%%%%%%%%%%%%%%%%%%%%%%%%%%%%%%%%%%%%%%%%%%%%%%%%%%%%%%%%%%%%%%
%%%%%%%%%%%%%%%%%%%%%%%%%%%%%%%%%%%%%%%%%%%%%%%%%%%%%%%%%%%%%%%%%%%%%%%%%%%%%%%
%

In the following, a scheme for the transformation of low-energy 
phonons in a given mode into an equal number of atoms in a different
atomic state with controlled energy and momentum based on doubly
detuned Raman transitions is presented.
Let us consider atoms which can be described by a three-level
($\Lambda$) system consisting of two (meta) stable states $\Psi_1$ and
$\Psi_2$ together with a third excited level $\Psi_3$ with the
energies $\omega_1<\omega_2<\omega_3$. 
This three-level system is illuminated by two optical Laser beams
which consist of many photons and can therefore be treated as rapidly 
oscillating classical fields described by the effective Rabi
frequencies $\Omega_1(t)$ and $\Omega_2(t)$.
Within the rotating wave and dipole approximation, the Lagrangian
reads ($\hbar=1$) 
\bea
\label{lambda}
L
=
i\Psi_1^*\dot\Psi_1+i\Psi_2^*\dot\Psi_2+i\Psi_3^*\dot\Psi_3
-\omega_1|\Psi_1^2|-\omega_2|\Psi_2^2|
\nn
-\omega_3|\Psi_3^2|
+[\Omega_1(t)\Psi_1^*\Psi_3+\Omega_2(t)\Psi_2^*\Psi_3+{\rm H.c.}]
\,.
\ea
The frequencies of the two doubly detuned Laser beams are chosen
according to (see Fig.~\ref{level})
\bea
\label{rabi}
\Omega_1(t)
&=&
\Omega_1\exp\{i(\omega_3-\omega_1+\Delta)t\}
\,,
\nn
\Omega_2(t)
&=&
\Omega_2\exp\{i(\omega_3-\omega_2+\Delta+\delta)t\}
\,,
\ea
with a large detuning $\Delta$ and a small detuning $\delta$ 
(which will later determine the phonon energy).
Introducing the slowly varying variables $\psi_1$, $\psi_2$, and
$\psi_3$ via 
$\Psi_1(t)=\psi_1(t)\exp\{-i\omega_1t\}$, 
$\Psi_2(t)=\psi_2(t)\exp\{-i\omega_2t\}$, and 
$\psi_3(t)\exp\{-i(\omega_3+\Delta)t\}$, we may solve 
the equation for the upper level~$\psi_3$ approximately for large
detuning~$\Delta$ via  
$\psi_3=-(\Omega_1^*\psi_1+\Omega_2^*e^{-i\delta t}\psi_2)
/\Delta+\ord(1/\Delta^2)$. 
%i.e., within the adiabatic approximation $|\dot\psi_3|\ll\Delta$.
%
Insertion into Eq.~(\ref{lambda}) yields the Lagrangian for the
remaining two levels in the adiabatic approximation 
$|\dot\psi_3|\ll\Delta$ 
\bea
\label{effective}
L_{\rm eff}
&=&
i\psi_1^*\dot\psi_1+i\psi_2^*\dot\psi_2
-\frac{|\Omega_1^2|}{\Delta}\,|\psi_1^2|
-\frac{|\Omega_2^2|}{\Delta}\,|\psi_2^2|
\nn
&&
-\left[
\frac{\Omega_1\Omega_2^*}{\Delta}\,e^{-i\delta t}\psi_1^*\psi_2
+{\rm H.c.}\right]
\,.
\ea
Assuming $|\Omega_1|=|\Omega_2|=\Omega$, both levels acquire the same
additional shift $\Omega^2/\Delta$; otherwise we would obtain an
effective detuning $\delta\to\delta'$ shifted by
$(|\Omega_1^2|-|\Omega_2^2|)/\Delta$. 

%%%%%%%%%%%%%%%%%%%%%%%%%%%%%%%%%%%%%%%%%%%%%%%%%%%%%%%%%%%%%%%%%%%%%%%%%%%%%%%
%%%%%%%%%%%%%%%%%%%%%%%%%%%%%%%%%%%%%%%%%%%%%%%%%%%%%%%%%%%%%%%%%%%%%%%%%%%%%%%
%{\em Many-particle formulation}\quad
%%%%%%%%%%%%%%%%%%%%%%%%%%%%%%%%%%%%%%%%%%%%%%%%%%%%%%%%%%%%%%%%%%%%%%%%%%%%%%%
%%%%%%%%%%%%%%%%%%%%%%%%%%%%%%%%%%%%%%%%%%%%%%%%%%%%%%%%%%%%%%%%%%%%%%%%%%%%%%%
%
An ideal quantum gas containing many of these atoms with mass $m$ can 
be described by the many-particle field operator $\hat\psi_r$ with the
dynamics (Heisenberg picture)
\bea
\label{many}
i\frac{\partial}{\partial t}\hat\psi_r
=
\left(-\frac{\na^2}{2m}+V_r\right)\hat\psi_r 
+
\Xi_{rs}\hat\psi_{s}
\,,
\ea
where $r,s=1,2$ are labels for the remaining two levels and 
$V_r$ the corresponding potentials.
The anti-hermitian space-time dependent transition amplitude 
$\Xi_{12}(t,x)=\exp\{-i\delta t+i\f{\kappa}\cdot\f{r}\}\Omega^2/\Delta$ 
represents the mode-coupling in Eq.~(\ref{effective}), where
$\f{\kappa}$ arises from a small angle between the Raman beams 
and the resulting wavenumber mismatch 
$\f{\kappa}=\f{k}_1^{\rm Laser}-\f{k}_2^{\rm Laser}$. 

An expansion into single-particle energy-eigenstates 
\bea
\label{single}
\hat\psi_s(t,\f{r})=\sum\limits_{\alpha}\hat a_{s\alpha}(t) 
f_{s\alpha}(\f{r})\exp\{-iE_{s\alpha}t\}
\,,
\ea
diagonalizes Eq.~(\ref{many}) apart from the transitions, which are
(in the rotating wave approximation) only relevant for 
$E_{1,\alpha}-E_{2,\beta}=\delta$ (energy conservation) 
and if the spatial matrix element 
$\bra{f_{1\alpha}}\hat\Xi_{12}\ket{f_{2\beta}}$ is large enough. 
For nearly homogeneous potentials $V_r\approx\rm const$, the
eigenfunctions are plane waves $\alpha\to\f{k}$ with 
$E_{r,\fk{k}}=\f{k}_r^2/(2m)+V_r$ and the latter condition represents
momentum conservation $\f{\kappa}=\f{k}_1-\f{k}_2$.
Hence, for a given frequency and wavenumber mismatch of the Lasers 
$(\delta,\f{\kappa})$, these energy and momentum conservation
conditions determine $\f{k}_1$ and $\f{k}_2$ up to a contribution
perpendicular to $\f{\kappa}$.
For effectively one-dimensional condensates, therefore, we can address 
single modes $\f{k}_1=k_1\f{e}_x$ and $\f{k}_2=k_2\f{e}_x$ by
adjusting the Lasers.

Now let us consider the following gedanken experiment:
Initially all atoms are in the state $r=1$ and form a nearly
homogeneous and (quasi) one-dimensional condensate, which is not in
its ground state but contains a single phonon with a given 
wavenumber~$\f{k}_{\rm p}=k_{\rm p}\f{e}_x$.  
In contrast to Eq.~(\ref{many}), this requires a non-vanishing
coupling $g$.
However, if we switch off this interaction $g$ adiabatically 
(e.g., via a Feshbach resonance), the system stays in this first
excited state and finally contains a single atom with the
momentum~$k_{\rm p}$ of the original phonon.   
After applying a Raman $\pi$-pulse 
(with the duration $T=\pi\Delta/\Omega^2$) adapted to this wavenumber,
e.g., $\kappa=k_{\rm p}$ and $\delta=k_{\rm p}^2/(2m)+V_1-V_2$,
exactly this single atom will be transferred to the other state $r=2$,
while all the condensate atoms are not affected  
(assuming that rotating wave approximation applies).

If we can separate the two species $r=1$ and $r=2$ or address them
individually, the number of atoms in the state $r=2$ can be counted 
via fluorescence measurements \cite{counting,raizen} and yields 
(in the ideal case) the number of phonons in a given mode~$k_{\rm p}$
present initially, i.e., one.
For example, a beam with a frequency just between the resonances of
the two species $r=1$ and $r=2$ is repulsive for one component and
attractive for the other one and could be used as optical tweezers. 

%%%%%%%%%%%%%%%%%%%%%%%%%%%%%%%%%%%%%%%%%%%%%%%%%%%%%%%%%%%%%%%%%%%%%%%%%%%%%%%
%%%%%%%%%%%%%%%%%%%%%%%%%%%%%%%%%%%%%%%%%%%%%%%%%%%%%%%%%%%%%%%%%%%%%%%%%%%%%%%
%{\em Mean-field formulation}\quad
%%%%%%%%%%%%%%%%%%%%%%%%%%%%%%%%%%%%%%%%%%%%%%%%%%%%%%%%%%%%%%%%%%%%%%%%%%%%%%%
%%%%%%%%%%%%%%%%%%%%%%%%%%%%%%%%%%%%%%%%%%%%%%%%%%%%%%%%%%%%%%%%%%%%%%%%%%%%%%%
%
With fixed momentum~$k_{\rm p}$ (homogeneous condensate), the energy
gap between the ground state and the one-particle excited state
decreases with diminishing coupling $g$ in view of the Bogoliubov
dispersion relation
$\omega^2(\f{k})=g\varrho\f{k}^2/m+\f{k}^4/(2m)^2$.
Hence, let us study the application of the Raman transitions in the
presence of a non-vanishing coupling $g$ 
(respecting the altered dispersion relation).
With interactions, the field operator $\hat\psi_r$ of  
the two levels $r=1,2$ obeys the equation of motion
\bea
\label{bec}
i\frac{\partial}{\partial t}\hat\psi_r
=
\left(-\frac{\na^2}{2m}+V_r+
g_{rs}\hat\psi^\dagger_{s}\hat\psi_{s}\right)\hat\psi_r
+
\Xi_{rs}\hat\psi_{s}
\,.
\ea
For simplicity, we assume equal one-particle trapping potentials
$V_1=V_2=V$ and coupling constants $g_{11}=g_{22}=g_{12}=g_{21}=g$ 
for the two species (otherwise we would again obtain an effective
detuning $\delta\to\delta'$). 
Initially, all the atoms (and hence also the condensate) are 
in the state $r=1$, which facilitates the mean-field expansion  
\bea
\label{mean}
\hat\psi_r
=
\left(
\begin{array}{c}
\hat\psi_1
\\
\hat\psi_2
\end{array}
\right)
=
\left(
\begin{array}{c}
\psi_c+\hat\chi
\\
\hat\zeta
\end{array}
\right)
\frac{\hat A}{\sqrt{N}}
+\ord(1/\sqrt{N})
\,,
\ea
with the condensate wave-function $\psi_c$ and the one-particle
excitations $\hat\chi$ and $\hat\zeta$.
The operator $\hat N=\hat A^\dagger\hat A$ counts the total number of 
particles.

In complete analogy to the previous example, the condensate $\psi_c$
is not affected by the Raman transitions for $\delta>0$ 
(assuming that the rotating wave approximation applies) and hence the
dynamics of the excitations read 
\bea
\label{transitions}
i\frac{\partial\hat\chi}{\partial t}
&=&
\left(-\frac{\na^2}{2m}+V+
2g\,|\psi_c^2|\right)\hat\chi+g\psi_c^2\hat\chi^\dagger
+\Xi\hat\zeta
\,,
\nn
i\frac{\partial\hat\zeta}{\partial t}
&=&
\left(-\frac{\na^2}{2m}+V+
g\,|\psi_c^2|\right)\hat\zeta
+\Xi^*\hat\chi
\,,
\ea
with 
$\Xi=\Xi_{12}=\exp\{-i\delta t+i\f{\kappa}\cdot\f{r}\}\Omega^2/\Delta$ 
as before. 
Assuming a nearly homogeneous condensate with   
$V+g\,|\psi_c^2|=\mu=\rm const$, a normal mode expansion yields 
\bea
\label{normal}
i\frac{\partial}{\partial t}\hat\zeta_{\fk{k}+\fk{\kappa}}
=
\left(\frac{(\f{k}+\f{\kappa})^2}{2m}+\mu\right)
\hat\zeta_{\fk{k}+\fk{\kappa}}
+\frac{\Omega^2}{\Delta}\,e^{i\delta t}\hat\chi_{\fk{k}}
\,.
\ea
The atomic one-particle excitation operator $\hat\chi_{\fk{k}}$ can be
decomposed into phonon creation and annihilation operators  
$\hat a^\dagger_{\fk{k}}$ and $\hat a_{\fk{k}}$, respectively 
($m=1$ for brevity)
\bea
\label{creation}
\hat\chi_{\boldsymbol k}
=
e^{-i\mu t}
\sqrt{\frac{{\boldsymbol k}^2}{2\omega_{\boldsymbol k}}}
\left[
\left(\frac12-\frac{\omega_{\boldsymbol k}}{{\boldsymbol k}^2}\right)
\hat a_{\boldsymbol k}^\dagger
+
\left(\frac12+\frac{\omega_{\boldsymbol k}}{{\boldsymbol k}^2}\right)
\hat a_{\boldsymbol k}
\right].
\ea
Inserting the time-dependences 
$\hat\zeta_{\fk{k}}(t)=\hat\zeta_{\fk{k}}
e^{-i[\mu+\fk{k}^2/(2m)]t}$
and
$\hat a_{\fk{k}}(t)=\hat a_{\fk{k}}e^{-i\omega_{\fk{k}}t}$
as well as applying the rotating wave approximation, only the second
term survives  
\bea
\label{phonon}
i\frac{\partial}{\partial t}\hat\zeta_{\fk{k}+\fk{\kappa}}
=
\frac{\Omega^2}{\Delta}\,
\sqrt{\frac{{\boldsymbol k}^2}{2\omega_{\boldsymbol k}}}
\left(\frac12+\frac{\omega_{\boldsymbol k}}{{\boldsymbol k}^2}\right)
\hat a_{\boldsymbol k}
\,,
\ea
and we obtain the expected resonance condition 
\bea
\label{resonance}
\delta=\omega_{\boldsymbol k}-\frac{(\f{k}+\f{\kappa})^2}{2m}
\,,
\ea
which implies energy conservation.
Of course, for $\f{\kappa}^2,\f{k}^2\gg1/\xi^2$, we reproduce the 
previous result~(\ref{effective}). 
Far below the healing length
$\f{\kappa}^2,\f{k}^2\ll1/\xi^2=mg\varrho$, i.e., in the phonon
regime, Eq.~(\ref{resonance}) simplifies to 
$\delta=\omega_{\boldsymbol k}$ and we get 
$i\partial\hat\zeta_{\fk{k}+\fk{\kappa}}/\partial t=
\sqrt{\mu/\delta}\,\hat a_{\fk{k}}\,\Omega^2/\Delta$.
Note that the pre-factor $\sqrt{\mu/\delta}$ is a consequence of the
interactions and illustrates the difference between phonons 
(``dressed'' atoms) and free-particle excitations as in
Eq.~(\ref{effective}).   

The effective interaction Hamiltonian from Eq.~(\ref{phonon}), 
\bea
\label{Hamiltonian}
\hat H_{\rm int}=
\frac{\Omega^2}{\Delta}\,
\sqrt{\frac{{\boldsymbol k}^2}{2\omega_{\boldsymbol k}}}
\left(\frac12+\frac{\omega_{\boldsymbol k}}{{\boldsymbol k}^2}\right)
\left(
\hat\zeta_{\fk{k}+\fk{\kappa}}^\dagger\hat a_{\fk{k}}+
\hat\zeta_{\fk{k}+\fk{\kappa}}\hat a_{\fk{k}}^\dagger
\right),
\ea
has the following intuitive interpretation:
Due to the detuning of the Raman beams, the missing energy~$\delta$
prohibits transitions from the multi-particle ground state of the
condensate (which has a sharp and well-defined energy) in component
$r=1$ to the other level $r=2$ and must be compensated by
absorbing a phonon with this (for $\f{k}=-\f{\kappa}$) or an even
higher energy $\omega_{\boldsymbol k}\geq\delta$.

If there are $n$ phonons to annihilate ($\hat a_{\fk{k}}$), $n$ atoms
can be transferred to the $r=2$ state 
($\hat\zeta_{\fk{k}+\fk{\kappa}}^\dagger$) such that the final number
of these transferred atoms measures the initial number of phonons.
Vice versa, if the component 2 is not empty initially, the Raman beams 
transfer atoms ($\hat\zeta_{\fk{k}+\fk{\kappa}}$) from the state 2 to
the level 1 with simultaneous emission of an equal number of phonons
($\hat a_{\fk{k}}^\dagger$).  

The energy-momentum balance (\ref{resonance}) is a bit more
complicated than in the previous case without interactions, but
exhibits a similar direction-degeneracy, which can again be eliminated
by considering effectively one-dimensional condensates.
In the phonon limit ($\lambda\gg\xi$ and $\delta\ll\mu$), we obtain a
unique solution for the phonon energy 
$\omega_{\boldsymbol k}\approx\delta$ which allows us to address
single modes with suitably tuned Lasers.
If we choose $\delta$ and $\f{\kappa}$ to lie on the phonon dispersion 
curve $\delta=\omega(\f{\kappa})$, we annihilate one phonon with energy 
$\delta$ and momentum $\f{\kappa}$ and create one particle in the
component $r=2$ in the ground state.

%%%%%%%%%%%%%%%%%%%%%%%%%%%%%%%%%%%%%%%%%%%%%%%%%%%%%%%%%%%%%%%%%%%%%%%%%%%%%%%
%%%%%%%%%%%%%%%%%%%%%%%%%%%%%%%%%%%%%%%%%%%%%%%%%%%%%%%%%%%%%%%%%%%%%%%%%%%%%%%
%{\em Detection scheme}\quad
%%%%%%%%%%%%%%%%%%%%%%%%%%%%%%%%%%%%%%%%%%%%%%%%%%%%%%%%%%%%%%%%%%%%%%%%%%%%%%%
%%%%%%%%%%%%%%%%%%%%%%%%%%%%%%%%%%%%%%%%%%%%%%%%%%%%%%%%%%%%%%%%%%%%%%%%%%%%%%%
%
With sufficiently long pulses leading to a good energy resolution, 
it should be possible to ``see'' the discrete nature of the phonon
spectrum, i.e., to address single (or a few) phonon modes. 
In order to annihilate all phonons in the $r=1$ condensate with a
given energy/momentum and to transfer the same number of atoms to the
$r=2$ component, we apply an effective Raman $\pi$-pulse with the
duration [cf.~Eq.~(\ref{Hamiltonian})]
\bea
\label{duration}
T
=
\frac{\pi\Delta}{\Omega^2}\,
\sqrt{\frac{2\omega_{\boldsymbol k}}{{\boldsymbol k}^2}}
\left(\frac12+\frac{\omega_{\boldsymbol k}}{{\boldsymbol k}^2}\right)^{-1}
\approx
\frac{\pi\Delta}{\Omega^2}\,
\sqrt{\frac{\delta}{\mu}}
\,,
\ea
where the $\approx$ sign applies to the phonon limit. 

%%%%%%%%%%%%%%%%%%%%%%%%%%%%%%%%%%%%%%%%%%%%%%%%%%%%%%%%%%%%%%%%%%%%%%%%%%%%%%%
%%%%%%%%%%%%%%%%%%%%%%%%%%%%%%%%%%%%%%%%%%%%%%%%%%%%%%%%%%%%%%%%%%%%%%%%%%%%%%%
%{\em Experimental parameters}\quad
%%%%%%%%%%%%%%%%%%%%%%%%%%%%%%%%%%%%%%%%%%%%%%%%%%%%%%%%%%%%%%%%%%%%%%%%%%%%%%%
%%%%%%%%%%%%%%%%%%%%%%%%%%%%%%%%%%%%%%%%%%%%%%%%%%%%%%%%%%%%%%%%%%%%%%%%%%%%%%%
%
Of course, the approximations used in the presented derivations must
be checked for a potentially realistic set of experimental parameters.  
Let us assume a speed of sound of a few millimeters per second and  
a healing length around one micrometer. 
In this case, the wavelength~$\lambda$ of the phonons to be detected
would typically be several micrometers $1/\kappa=\ord(10\,\mu\rm m)$
and their frequency a few hundred Hertz $\delta=\ord(100\,\rm Hz)$. 

Using Lasers in the optical range $\ord(10^{15}\,\rm Hz)$, the large
detuning $\Delta$ depends on the atomic level structure and would be a
little bit below this value, say $\Delta=\ord(10^{13}-10^{14}\,\rm Hz)$.
With quite moderate Rabi frequencies $\Omega=\ord(10^4-10^7\,\rm Hz)$,
we can achieve an effective Raman transition rate
$\sqrt{\mu/\delta}\;\Omega^2/\Delta$ of a few tens of Hertz.
Consequently, the duration of the effective Raman $\pi$-pulse in
Eq.~(\ref{duration}) would be of the order of hundred milliseconds
$T=\ord(100\,\rm ms)$ leading to a energy resolution of circa ten
Hertz.  
In view of the aforementioned parameters, the assumptions and
approximations (e.g., the adiabaticity $\Delta\gg\Omega$)
used in the derivation are reasonably well justified. 
The major constraint is given by the energy resolution of the
effective Raman $\pi$-pulse peaked around 
$\delta=\ord(100\,\rm Hz)\pm\ord(10\,\rm Hz)$.
Apart from a few excitations (i.e., phonons), the beams illuminate
many atoms in the ground state (zero energy) and one has to make sure
that the probability of transferring an atom from the ground state of
the condensate in component $r=1$ into the state $r=2$ is small
enough. 
Thus the negative-frequency tail of the Fourier transform of
the pulse (which is peaked around $\delta$ in frequency space) must be 
suppressed accordingly \cite{blackman,raizen}.

%%%%%%%%%%%%%%%%%%%%%%%%%%%%%%%%%%%%%%%%%%%%%%%%%%%%%%%%%%%%%%%%%%%%%%%%%%%%%%%
%%%%%%%%%%%%%%%%%%%%%%%%%%%%%%%%%%%%%%%%%%%%%%%%%%%%%%%%%%%%%%%%%%%%%%%%%%%%%%%
%{\em Applications}\quad
%%%%%%%%%%%%%%%%%%%%%%%%%%%%%%%%%%%%%%%%%%%%%%%%%%%%%%%%%%%%%%%%%%%%%%%%%%%%%%%
%%%%%%%%%%%%%%%%%%%%%%%%%%%%%%%%%%%%%%%%%%%%%%%%%%%%%%%%%%%%%%%%%%%%%%%%%%%%%%%
%
With the ability of measuring a few low-energy phonons, it might
become possible to observe some of the exotic quantum effects
mentioned in the Introduction. 
The analogue of cosmological particle production is probably easier to
realize experimentally than Hawking radiation since it can be done
with a condensate at rest and a practically unlimited measurement
time (after varying $c_s$ via a Feshbach resonance, for example).
The same advantage applies to a small wiggling stirrer in the
condensate, which would act as a point-like non-inertial scatterer and
generate the analogue of moving-mirror radiation \cite{birrell}, which
can be interpreted as a signature of the Unruh effect.
In contrast, the detection of the Hawking radiation requires either a
flowing condensate or a motion of the horizon via a space-time 
dependent sound velocity $c_s(t,x)$, cf.~\cite{wave-guide}.
Apart from measuring this striking effect, these experiments may also 
shed light onto the trans-Planckian problem, i.e., impact of the
short-range physics on the long-wavelength Hawking radiation:
Even though the Hawking effect seems to be quite robust against
modifications of the dispersion relation at short wavelengths 
(such as the Bogoliubov dispersion in Bose-Einstein condensates) 
only very little is known about the impact of interactions, see, e.g.,
\cite{Universality}.  

%%%%%%%%%%%%%%%%%%%%%%%%%%%%%%%%%%%%%%%%%%%%%%%%%%%%%%%%%%%%%%%%%%%%%%%%%%%%%%%
\acknowledgments
%%%%%%%%%%%%%%%%%%%%%%%%%%%%%%%%%%%%%%%%%%%%%%%%%%%%%%%%%%%%%%%%%%%%%%%%%%%%%%%
%
%%%%%%%%%%%%%%%%%%%%%%%%%%%%%%%%%%%%%%%%%%%%%%%%%%%%%%%%%%%%%%%%%%%%%%%%%%%%%%%
%%%%%%%%%%%%%%%%%%%%%%%%%%%%%%%%%%%%%%%%%%%%%%%%%%%%%%%%%%%%%%%%%%%%%%%%%%%%%%%
{\em Acknowledgments}\quad
%%%%%%%%%%%%%%%%%%%%%%%%%%%%%%%%%%%%%%%%%%%%%%%%%%%%%%%%%%%%%%%%%%%%%%%%%%%%%%%
%%%%%%%%%%%%%%%%%%%%%%%%%%%%%%%%%%%%%%%%%%%%%%%%%%%%%%%%%%%%%%%%%%%%%%%%%%%%%%%
%
The idea to the presented detection scheme was developed together with
Mark Raizen \cite{raizen} and emerged during the workshop 
``Low dimensional Systems in Quantum Optics'' in September 2005 
at the Centro Internacional de Ciencias in Cuernavaca (Mexico), 
which was supported by the Alexander von Humboldt foundation.
This work was supported by the Emmy-Noether Programme of the German
Research Foundation (DFG) under grant No.~SCHU 1557/1-2.
Further support by the ESF-COSLAB and the EU-ULTI
programmes is also gratefully acknowledged.  

%%%%%%%%%%%%%%%%%%%%%%%%%%%%%%%%%%%%%%%%%%%%%%%%%%%%%%%%%%%%%%%%%%%%%%%%%%%%%%%
%%%%%%%%%%%%%%%%%%%%%%%%%%%%%%%%%%%%%%%%%%%%%%%%%%%%%%%%%%%%%%%%%%%%%%%%%%%%%%%
%%%%%%%%%%%%%%%%%%%%%%%%%%%%%%%%%%%%%%%%%%%%%%%%%%%%%%%%%%%%%%%%%%%%%%%%%%%%%%%
%%%%%%%%%%%%%%%%%%%%%%%%%%%%%%%%%%%%%%%%%%%%%%%%%%%%%%%%%%%%%%%%%%%%%%%%%%%%%%%

%%%%%%%%%%%%%%%%%%%%%%%%%%%%%%%%%%%%%%%%%%%%%%%%%%%%%%%%%%%%%%%%%%%%%%%%%%%%%%%
%%%%%%%%%%%%%%%%%%%%%%%%%%%%%%%%%%%%%%%%%%%%%%%%%%%%%%%%%%%%%%%%%%%%%%%%%%%%%%%
%%%%%%%%%%%%%%%%%%%%%%%%%%%%%%%%%%%%%%%%%%%%%%%%%%%%%%%%%%%%%%%%%%%%%%%%%%%%%%%
%%%%%%%%%%%%%%%%%%%%%%%%%%%%%%%%%%%%%%%%%%%%%%%%%%%%%%%%%%%%%%%%%%%%%%%%%%%%%%%
%%%%%%%%%%%%%%%%%%%%%%%%%%%%%%%%%%%%%%%%%%%%%%%%%%%%%%%%%%%%%%%%%%%%%%%%%%%%%%%
%%%%%%%%%%%%%%%%%%%%%%%%%%%%%%%%%%%%%%%%%%%%%%%%%%%%%%%%%%%%%%%%%%%%%%%%%%%%%%%
%%%%%%%%%%%%%%%%%%%%%%%%%%%%%%%%%%%%%%%%%%%%%%%%%%%%%%%%%%%%%%%%%%%%%%%%%%%%%%%
%%%%%%%%%%%%%%%%%%%%%%%%%%%%%%%%%%%%%%%%%%%%%%%%%%%%%%%%%%%%%%%%%%%%%%%%%%%%%%%
%%%%%%%%%%%%%%%%%%%%%%%%%%%%%%%%%%%%%%%%%%%%%%%%%%%%%%%%%%%%%%%%%%%%%%%%%%%%%%%
%%%%%%%%%%%%%%%%%%%%%%%%%%%%%%%%%%%%%%%%%%%%%%%%%%%%%%%%%%%%%%%%%%%%%%%%%%%%%%%
%%%%%%%%%%%%%%%%%%%%%%%%%%%%%%%%%%%%%%%%%%%%%%%%%%%%%%%%%%%%%%%%%%%%%%%%%%%%%%%
%%%%%%%%%%%%%%%%%%%%%%%%%%%%%%%%%%%%%%%%%%%%%%%%%%%%%%%%%%%%%%%%%%%%%%%%%%%%%%%
%%%%%%%%%%%%%%%%%%%%%%%%%%%%%%%%%%%%%%%%%%%%%%%%%%%%%%%%%%%%%%%%%%%%%%%%%%%%%%%
%%%%%%%%%%%%%%%%%%%%%%%%%%%%%%%%%%%%%%%%%%%%%%%%%%%%%%%%%%%%%%%%%%%%%%%%%%%%%%%
%%%%%%%%%%%%%%%%%%%%%%%%%%%%%%%%%%%%%%%%%%%%%%%%%%%%%%%%%%%%%%%%%%%%%%%%%%%%%%%
%%%%%%%%%%%%%%%%%%%%%%%%%%%%%%%%%%%%%%%%%%%%%%%%%%%%%%%%%%%%%%%%%%%%%%%%%%%%%%%
\end{document}